\documentclass[journal]{IEEEtran}
\usepackage{cite}
\usepackage{amsmath,amssymb,amsfonts}
\usepackage{algorithmic}
\usepackage{graphicx}
\usepackage{textcomp}
\usepackage{booktabs}
\usepackage{multirow}
\usepackage{array}

\usepackage{bm}
\makeatletter
\AtBeginDocument{\DeclareMathVersion{bold}
\SetSymbolFont{operators}{bold}{T1}{times}{b}{n}
\SetMathAlphabet{\mathrm}{bold}{T1}{times}{b}{n}
\SetMathAlphabet{\mathit}{bold}{T1}{times}{b}{it}
\SetMathAlphabet{\mathbf}{bold}{T1}{times}{b}{n}
\SetMathAlphabet{\mathtt}{bold}{OT1}{pcr}{b}{n}
\SetSymbolFont{symbols}{bold}{OMS}{cmsy}{b}{n}
\renewcommand\boldmath{\@nomath\boldmath\mathversion{bold}}}
\makeatother

\def\BibTeX{{\rm B\kern-.05em{\sc i\kern-.025em b}\kern-.08em
    T\kern-.1667em\lower.7ex\hbox{E}\kern-.125emX}}

\begin{document}
% \history{Date of publication xxxx 00, 0000, date of current version xxxx 00, 0000.}
% \doi{10.1109/ACCESS.2024.0429000}

\title{Watermarks Without Verification: AI Text Watermarking After the EU AI Act}

\author{%
Alexander Nemecek$^{\ast}$,
Vipin Chaudhary,
Erman Ayday
\\
Case Western Reserve University
\thanks{$^{\ast}$Correspondence: \texttt{ajn98@case.edu}}%
}

% \address[1]{Case Western Reserve University, Cleveland, OH 44106 USA}

% \author{\uppercase{First A. Author}\authorrefmark{1}, \IEEEmembership{Fellow, IEEE},
% \uppercase{Second B. Author}\authorrefmark{2}, AND \uppercase{Third C. Author},
% Jr.\authorrefmark{3},
% \IEEEmembership{Member, IEEE}}

% \markboth
% {Nemecek \headeretal: Watermarks Without Verification: AI Text Watermarking After the EU AI Act}
% {Nemecek \headeretal: Watermarks Without Verification: AI Text Watermarking After the EU AI Act}

% \corresp{Corresponding author: Alexander Nemecek (e-mail: ajn98@case.edu).}

\maketitle

\begin{abstract}
On August 2, 2026, the obligations of Article 50 of the EU AI Act took effect, requiring generative AI providers to mark the content their systems produce and ensure it can be detected as AI-generated. Days later, Anthropic disclosed that every Claude model released after that date embeds a watermark based on SynthID-Text in all generated text, enabled by default with no user opt-out; Google has deployed SynthID-Text in Gemini since 2024. Users objected that the watermark degrades quality, particularly for code, that it secretly encodes identifying information, and, in mutual contradiction, that it is easily removable and inescapable; the vendor answered with assurances of unchanged quality, no identifying information, and robustness to light editing. In this work, we argue that neither the objections nor the assurances can currently be verified and that this unverifiability, rather than watermarking itself, is the substantive governance failure. We sort the contested assertions by what it would take to settle each and evaluate the open-source SynthID-Text implementation on two open-weight models, because no public tool can test the deployed systems. On prose, the measured effect of the watermark does not exceed that of changing the sampling seed. On code, the cost is three points of correctness on one model and below measurement on the other, while detection remains near chance, a limitation of detectability rather than quality. The remaining gaps trace to withheld access or missing institutions and we map each to a requirement: release of matched outputs, configuration disclosure, accredited audits, a shared evaluation protocol, and interoperable detection.
\end{abstract}


% \titlepgskip=-21pt

\section{Introduction}\label{sec:introduction}

On August 2, 2026, the obligations of Article 50 of the EU AI Act took effect, requiring providers of generative AI systems to mark the content their systems produce in a machine-readable way and to ensure that it can be detected as AI-generated~\cite{euaiact2024}. By that date, approximately 190 organizations, including OpenAI, Anthropic, Meta, Google, and Microsoft, had signed the voluntary Code of Practice on Transparency of AI-Generated Content issued by the European Commission's AI Office~\cite{codeofpractice2026}. The Code recognizes watermarking as one acceptable marking technique and commits signatories to providing detection systems. 

Text watermarking is not new and it has already been deployed at scale. Google has embedded its SynthID-Text watermark in Gemini's consumer products since 2024 and has published the method along with an evaluation run on nearly twenty million live responses~\cite{dathathri2024scalable}. Anthropic's disclosure on its support pages in August 2026~\cite{anthropic2026marking}, expanded in a technical post on August 14~\cite{anthropic2026watermark}, brought the technology to broader public attention~\cite{techcrunch2026}. Anthropic stated that every Claude model released after August 2 would embed a watermark, based on the SynthID-Text approach, in all generated text; that the watermark would be enabled by default, with no option for users to disable it; and that a public detection API, an interface through which outside parties could check text for the mark, would follow~\cite{anthropic2026marking, anthropic2026watermark}.

Since the rest of the debate turns on how such a watermark works, we describe the mechanism briefly. A statistical text watermark of this kind operates at the moment the model chooses each word. Among the candidate words the model considers acceptable, it slightly favors those that a secret key singles out~\cite{kirchenbauer2023watermark, aaronson2023watermarking, dathathri2024scalable, nemecek-etal-2026-topic}. Nothing is added to the text and the text reads normally. A detector holding the same key, however, can count how often the favored choices appear and, over a long enough passage, judge whether the text came from the model~\cite{kirchenbauer2023watermark, dathathri2024scalable}. Because the mark is applied while the model samples each word, watermarks of this kind are called sampling-based~\cite{liu2024survey}. Where the model has few acceptable choices, as in code or short factual answers, there is little room to steer the choice and the signal is weaker~\cite{dathathri2024scalable, lee2024sweet}.

The public reaction to Anthropic's announcement was immediate. To our knowledge, it is the largest real-world test to date of how the public responds to generative AI watermarking. Within days, online discussion involving thousands of participants converged on a small set of objections~\cite{gizmodo2026revolt}. The most prominent was that watermarking generated code would unavoidably degrade it. Commentators characterized the watermark as an unacceptable adulteration of the generated text itself. The business press reported users canceling subscriptions in protest~\cite{businessinsider2026cancel}. A tool posted on GitHub that claimed to remove the watermark collected more than ten thousand stars, the platform's bookmark-style endorsement, within a week of the announcement, although no evidence of its effectiveness had been published~\cite{meyer2026remover, thenextweb2026removers, bleeping2026removers}. 

The users' positions cluster around four claims: (Claim 1) the watermark degrades the quality of the output; (Claim 2) the watermark secretly encodes identifying information, enabling surveillance; (Claim 3) the watermark is easily removable and therefore useless; and (Claim 4) the watermark is effectively non-removable and therefore inescapable. We note at the outset that some of these claims contradict one another; a watermark cannot be both easily removable and inescapable. We return to this tension in Section~\ref{sec:conflict}.

Anthropic's technical communication makes correspondingly strong assurances in response: (Assurance 1) the watermark is imperceptible to readers; (Assurance 2) it has no practical impact on the quality or content of the output; (Assurance 3) no hidden characters are inserted into the text, and generating watermarked text consumes no additional tokens, the units in which model output is measured and billed; (Assurance 4) the watermark carries no identifying information and cannot be traced to a specific person, organization, or conversation; and (Assurance 5) the watermark survives light editing, although a complete rewrite will remove it~\cite{anthropic2026watermark, anthropic2026marking}. Anthropic is not the first provider to deploy text watermarking at scale and these assurances match what Google has published about its own deployment~\cite{dathathri2024scalable, synthid2024docs}. For neither deployment, however, can anyone outside the company currently verify them.

This unverifiability is what motivates the present article. The users' claims contradict a body of measured results on sampling-based watermarking~\cite{kirchenbauer2023watermark, kuditipudi2023robust, dathathri2024scalable, kirchenbauer2024reliability}, and those results were largely absent from the public discussion. The vendor's assurances are broadly consistent with the same results, but the results concern the general method, which we call the watermark family, evaluated on models whose weights are publicly downloadable (open-weight models), rather than the proprietary system actually deployed. The deployed configuration cannot be inspected by anyone outside the company. An independent audit would need to cover the sampling parameters that control how strongly the watermark is applied; the entropy thresholds below which it is switched off, since highly predictable text leaves too little room to embed a signal~\cite{dathathri2024scalable, lee2024sweet}; how the secret keys are managed; how the detector is calibrated; and how the watermark behaves across languages and content types. There is no side-by-side comparison of watermarked and unwatermarked output from the same model, and the planned detection API will be governed solely by the vendor~\cite{anthropic2026watermark}. In the absence of independent verification, it is rational for users to fill the gap with worst-case assumptions. The information vacuum is filled with rumors, mutually inconsistent objections, and removal tools that no one has shown to work~\cite{bleeping2026removers}. The result is a verification problem that the public experiences as a trust problem.

This article makes four contributions. First, we organize the contested claims and assurances into a taxonomy sorted by what it would take to settle each one: some can be tested with public tools, some only with the vendor's cooperation, and some cannot be verified at all until standards and audit institutions exist (Section~\ref{sec:conflict}). Second, after situating the debate in the current literature (Section~\ref{sec:literature}), we report a targeted evaluation of the watermark family that providers report deploying (SynthID-Text), covering prose quality and code generation (Section~\ref{sec:eval}). On prose, the measured effect of the watermark does not exceed that of changing the sampling seed; on code, the cost is at most three points of correctness on one model and below measurement on the other, while detection remains near chance. Third, we analyze the claims that cannot currently be verified and argue that this unverifiability, rather than watermarking itself, is the substantive governance failure (Section~\ref{sec:problem}). Fourth, we map each unverifiable claim to a concrete institutional requirement, such as standardized evaluation protocols and independent audit access (Section~\ref{sec:moving_forward}). Finally, we close with an outlook (Section~\ref{sec:outlook}): the current episode is a preview of what is to come, not an anomaly. Whether watermarking becomes a trusted way of establishing where content came from or an empty compliance ritual will depend on whether institutions capable of verifying vendors' claims arrive before public trust erodes further.

\section{Claims in Conflict}\label{sec:conflict}

The assertions discussed in this section are taken from provider documentation, press coverage, and widely read online discussion threads following the August 2026 deployment. We distill each position into its strongest recurring form. This is an illustrative reconstruction, not a systematic survey of everything that was said, and we make no claims about how common any position is. Although the Claude deployment triggered the current debate, few of the assertions are new. Google made equivalent quality and imperceptibility claims when it deployed SynthID-Text in 2024~\cite{dathathri2024scalable, synthid2024docs}. OpenAI raised several of the users' objections itself in 2024, when it explained why it had built a text watermark and chosen not to release it~\cite{openai2024provenance, wsj2024openaiwatermark}. Suspicion of covert marking predates the Claude deployment entirely: in 2025, invisible Unicode characters were discovered in ChatGPT output, which OpenAI attributed to a byproduct of training~\cite{rumi2025watermarks}. In addition to the four claims and five assurances defined in Section~\ref{sec:introduction}, the taxonomy therefore includes one limitation that the vendor itself acknowledges and two concerns raised by another provider, since a taxonomy restricted to skeptics and a single vendor would misrepresent the debate.

Table~\ref{tab:claims} sorts each assertion by what it would take to settle it. An assertion is publicly testable, category (a), when publicly available tools and models suffice to produce decisive evidence. It requires vendor cooperation, category (b), when settling it depends on access that only the vendor can grant, such as matched pairs of outputs generated with and without the watermark, or a detector that reports a confidence score rather than a yes/no answer. It requires new institutions, category (c), when no external experiment can settle it and verification depends on bodies that do not yet exist, such as an independent audit of how the secret keys are managed. One further distinction runs through the table: evidence about a watermark family is not evidence about a deployment. SynthID-Text is published, open-sourced, and measurable~\cite{dathathri2024scalable, synthidtext2024code}; the deployed configurations are none of these things. Several assertions therefore occupy two categories at once, and much of the public dispute stems from treating evidence about the family as though it settled questions about the deployment, and vice versa.

\begin{table*}[!t]
\centering
\footnotesize
\begin{tabular}{@{}>{\raggedright\arraybackslash}p{0.1\textwidth}>{\raggedright\arraybackslash}p{0.25\textwidth}>{\raggedright\arraybackslash}p{0.12\textwidth}>{\raggedright\arraybackslash}p{0.17\textwidth}>{\raggedright\arraybackslash}p{0.25\textwidth}@{}}
\toprule
\textbf{Property} & \textbf{Assertion and source} & \textbf{Category} & \textbf{Settling evidence} & \textbf{Current status} \\
\midrule
Quality & Claim 1 (users). Watermarking degrades output quality. & (a) family; \newline (b) deployment & Matched watermark-on and watermark-off outputs under a fixed quality protocol. & Family-level evidence finds negligible degradation for prose; no deployment has released matched outputs (Section~\ref{sec:eval}). \\
 & Assurance 2 (Google 2024~\cite{dathathri2024scalable}; Anthropic 2026~\cite{anthropic2026watermark}). No practical impact on quality or content. & (a) family; \newline (b) deployment & Same as Claim 1. & Family-level results find negligible degradation; Google reports preference parity at production scale from its own deployment, but the evaluation was run by the vendor and reported only in aggregate; no equivalent evidence exists for Anthropic. \\
\midrule
Mechanics & Assurance 1 (Google 2024~\cite{dathathri2024scalable}; Anthropic 2026~\cite{anthropic2026watermark}). The watermark is imperceptible to readers. & (a) family; \newline (b) deployment & Blinded pairwise human preference. & Google's production evaluation found no user-preference difference; the result is vendor-run and unreplicated; no equivalent exists for Anthropic. \\
 & Assurance 3 (Anthropic 2026~\cite{anthropic2026watermark}). No hidden characters and no additional tokens. & (a) deployment & Character-level inspection and token accounting through public interfaces. & Community inspection to date is consistent with the assurance~\cite{pillitteri2026tested}. \\
\midrule
Traceability & Claim 2 (users). The watermark covertly encodes identifying information, enabling surveillance. & (c) & Independent audit of key management and of how keys are assigned. & The published family embeds no information about the user~\cite{dathathri2024scalable}; how the deployment assigns keys is not externally observable. \\
 & Assurance 4 (Anthropic 2026~\cite{anthropic2026watermark}; implicit in Google's published description of its method~\cite{dathathri2024scalable}). No identifying information and no tracing to a person, organization, or conversation. & (c) & Same as Claim 2. & A design claim and the mirror image of Claim 2; unverifiable without audit access. \\
\midrule
Removability & Claim 3 (users; overlapping OpenAI's 2024 withholding rationale~\cite{openai2024provenance}). Easily removable and therefore useless. & (a) family; \newline (b) deployment & A standardized suite of editing and paraphrasing tests run against the deployed detector. & Paraphrase and translation substantially reduce family-level detectability~\cite{krishna2023paraphrasing, he2024can}; the mark largely survives light edits~\cite{kuditipudi2023robust}. \\
 & Claim 4 (users). Effectively non-removable and therefore inescapable. & (a) family; \newline (b) deployment & Same as Claim 3. & Contradicted by the same robustness results that limit Claim 3. \\
 & Assurance 5 (Anthropic 2026~\cite{anthropic2026watermark}; Google 2024 acknowledges paraphrase and translation weakness~\cite{dathathri2024scalable}). Survives light editing; a full rewrite removes it. & (a) family; \newline (b) deployment & Same as Claim 3. & Consistent with family-level results; deployment thresholds unknown. \\
\midrule
Misattribution & Limitation 1 (Anthropic-acknowledged~\cite{anthropic2026marking, anthropic2026watermark}). The mark records model involvement rather than authorship, so human-written text edited by the model also carries it. & (b); (a) once a scored detector exists & Detector behavior on model-generated text versus human text lightly edited by the model. & Acknowledged by the vendor; the real-world misattribution rate is unknown. \\
 & Concern 1 (OpenAI 2024~\cite{openai2024provenance}). At scale, even a low false-positive rate produces large volumes of wrongly flagged text. & (b); (c) & Disclosed detector calibration and reporting of error rates at population scale. & Voiced by a provider as grounds for withholding; no deployment has published calibration. \\
\midrule
Disparate impact & Concern 2 (OpenAI 2024~\cite{openai2024provenance}). Detection burdens fall unevenly, in particular on non-native speakers who rely more heavily on AI writing assistance. & (a) family;  \newline (b) deployment & Disaggregated quality and detectability evaluation across languages and user groups. & Voiced by a provider; no deployment reports disaggregated results (Section~\ref{sec:problem}). \\
\bottomrule
\end{tabular}
\vspace{1em}
\caption{Contested assertions in the text watermarking debate, grouped by property. Categories are (a) publicly testable, (b) testable only with vendor cooperation, and (c) unverifiable absent standards and audit institutions.}
\label{tab:claims}
\end{table*}

The quality dispute between Claim 1 and Assurance 2, together with the perceptibility question of Assurance 1, can be evaluated today at the family level, using open-weight models and the open-source SynthID-Text implementation~\cite{synthid2024docs} (Section~\ref{sec:eval}). Assurance 3 stands apart because it is the only assertion testable on the deployment itself: hidden characters can be found by inspecting the text character by character, and token counts can be read from the public interface. This row matters for the credibility of the taxonomy. In the one case where outside verification is possible, the vendor's account holds~\cite{pillitteri2026tested}, and the taxonomy functions as an instrument instead of an indictment.

The rows that require vendor cooperation share a single missing resource: access. Matched pairs of outputs from the deployed model, generated with and without the watermark, would settle the quality and perceptibility questions for Claude specifically. A detection interface that reports a confidence score rather than a yes/no answer would allow the removability assertions to be tested against the system that is actually deployed. Limitation 1~\cite{anthropic2026marking, anthropic2026watermark} sits in this category for now, since the rate at which human-written text edited by the model is flagged cannot be measured until a detector is available; we return to its implications for governance in Section~\ref{sec:problem}. Each form of access named here reappears as a requirement in Section~\ref{sec:moving_forward}.

The traceability pair occupies category (c) alone, and the two assertions are mirror images of one another. Claim 2 asserts a capability that cannot be demonstrated from outside, and Assurance 4 denies a capability in a way that cannot be confirmed from outside. The crux is how the secret keys are assigned. The published method embeds no information about the user in the text~\cite{dathathri2024scalable}. But if a deployment used a different key for each customer or organization, the operator could identify the source of a text simply by checking which key matches, without adding anything to the text at all~\cite{yoo2024advancing, fernandez2023three}, and no external test can distinguish the two configurations. The vendor and its critics are therefore stuck in symmetric positions: neither can prove its case, and no external experiment can resolve the question.

Claims 3 and 4 cannot both be true. A signal that is easily removable cannot also be inescapable, yet both circulate in the same threads, invoked as the argument requires. The charge that the watermark is removable supports the conclusion that it is useless; the charge that it is permanent supports the conclusion that it enables surveillance. Published research on how well watermarks survive editing contradicts both claims. Rewriting a text in different words (a paraphrase attack) does remove the mark, which refutes permanence~\cite{sadasivan2023can, krishna2023paraphrasing}, while the mark's persistence under light editing refutes the claim that it is easily removed~\cite{kuditipudi2023robust, kirchenbauer2024reliability, nemecek-etal-2026-topic}. That mutually exclusive objections circulate side by side is a sign of a debate proceeding without shared evidence, which is the gap the taxonomy is meant to fill. The category (a) rows proceed to Section~\ref{sec:eval}, and the remainder to Section~\ref{sec:problem}.

\section{What the Literature Already Says}\label{sec:literature}

This section collects what is already established about sampling-based watermarking and where the established results stop, using the properties of Table~\ref{tab:claims} as the organizing frame. Several surveys cover the field as a whole~\cite{liu2024survey}; our aim is narrower in order to show which parts of the August 2026 debate re-litigate settled questions and which do not.

\subsection{Distortion-Based and Distortion-Free Watermarks}
\label{sec:literature:families}

Section~\ref{sec:introduction} described a watermarking mechanism in which the model slightly favors tokens selected by a secret key, and a detector holding that key counts how often those tokens appear. A token is the unit a language model actually produces, a word or fragment of a word from a fixed vocabulary. At each step the model computes a probability distribution over that vocabulary, the next-token distribution, and samples the next token from it. A watermark intervenes at this sampling step. One further distinction is whether the watermark changes the next-token distribution or only the randomness used to sample from it. The published schemes fall into two families.

In the first family, the watermark biases the token distribution. At each step, the key partitions the vocabulary into a green list of favored tokens and a red list of the rest, and the logits, the model's raw scores for each token, are raised by a fixed amount for the green tokens before the next token is sampled~\cite{kirchenbauer2023watermark}. A variant fixes the partition once rather than re-deriving it at each step, trading some resistance to reverse engineering for robustness to editing~\cite{zhao2023provable}. Watermarked text is sampled from a slightly different distribution than the model would otherwise produce, and the cost, although modest, can be measured, usually as an increase in perplexity. The literature calls these distortion-based schemes.

In the second family, the next-token distribution is left untouched, and the watermark is carried by the randomness used to sample from it. The key fixes the pseudorandom values that decide which of several equally acceptable tokens is drawn, so that any single response is distributed as it would have been without the watermark~\cite{aaronson2023watermarking, kuditipudi2023robust}. The strongest version of the idea makes watermarked and unwatermarked text computationally indistinguishable~\cite{christ2024undetectable}. There is no quality cost to measure, by construction. The signal depends entirely on the entropy of the next-token distribution, that is, on how much choice the model had. Where the next token is nearly determined, as in code, structured output, or short factual answers, there is little for the key to decide, and the watermark is faint or absent. The literature calls these distortion-free schemes. The guarantee concerns a single response: with one fixed key, repeated responses to the same prompt become more alike than they otherwise would. This is a loss of diversity across responses rather than a change to any one of them, and implementations mitigate it by varying the seed.

SynthID-Text belongs to the second family, with an adjustable strength. At each step it samples several candidate tokens from the next-token distribution, assigns each a pseudorandom g-value derived from the key and the preceding tokens, and runs the candidates through a tournament in which the candidate with the higher g-value advances; the number of tournament layers sets how strongly the key's preference is imposed~\cite{dathathri2024scalable}. In its non-distortionary configuration it leaves individual responses unchanged; in its distortionary configuration, however, it behaves more like the first family and yields a stronger signal. The detector recomputes the g-values for each token of a text and scores the text by their mean. It is the only scheme of either family deployed at consumer scale, its authors reported a live comparison across nearly twenty million Gemini responses, and a reference implementation has been released and integrated into a widely used open-source model library~\cite{synthidtext2024code}. That combination, deployed, published, and open, is what makes the family-level evaluation of Section~\ref{sec:eval} possible. Which configuration Anthropic runs, at what strength, and with what entropy cutoff, if any, below which tokens are left unwatermarked~\cite{lee2024sweet}, is a deployment question.

Two further variants bear on two rows of Table~\ref{tab:claims}. A multi-bit watermark carries a payload, a short message such as a model version or an account identifier, by letting the message select which key or which partition of the vocabulary is used~\cite{yoo2024advancing, fernandez2023three}. The published SynthID-Text scheme is zero-bit since it carries no payload and its detector returns only a score. The existence of multi-bit schemes means that Assurance 4 describes a design choice rather than a technical limit, which is why it sits in category (c). Separately, detection normally requires the same secret key used for embedding, but publicly detectable schemes exist in which anyone holding a public key can verify the mark without being able to forge it~\cite{fairoze2023publicly}. Who is able to run a detector is likewise a design choice.

\begin{figure*}[!t]
\centering
\includegraphics[width=\textwidth]{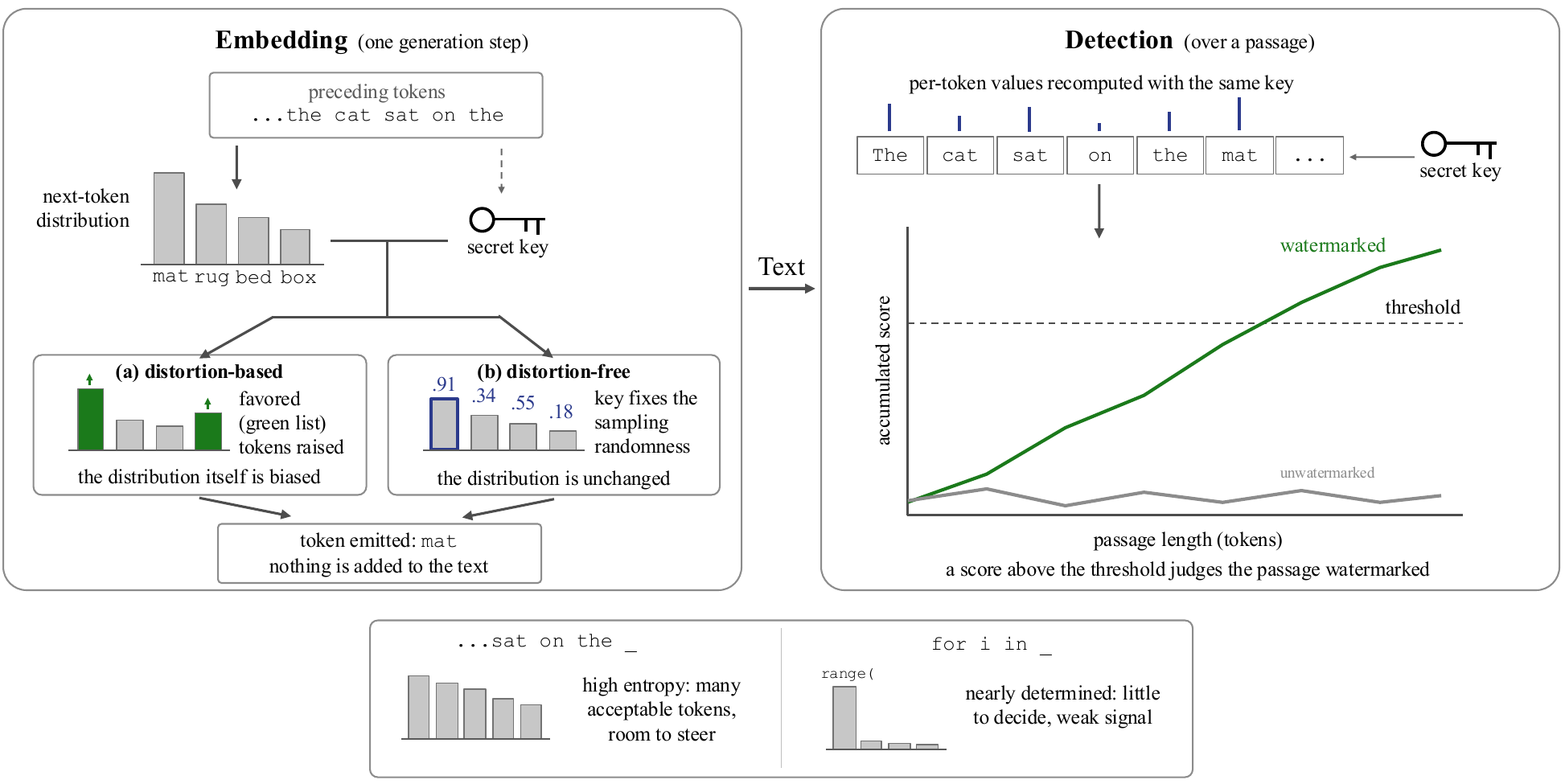}
\caption{How a sampling-based text watermark is embedded and detected. Left: at each step the model computes its next-token distribution as usual; a secret key and the preceding tokens determine which tokens are favored, either by biasing the distribution (distortion-based) or by fixing the randomness used to sample from it (distortion-free). Right: a detector holding the same key recomputes the per-token values for a passage and accumulates a score, which grows with the length of the passage and with the entropy of each step. Where the next token is nearly determined (inset), the key has little to decide, and the signal is weak.}
\label{fig:mechanism}
\end{figure*}

Finally, much of the public's intuition about AI-text detection comes from a different kind of tool. Post-hoc detectors examine unmarked text for statistical or stylistic signs of machine authorship, using either a classifier trained on human and machine text or statistics computed from a model's own token probabilities~\cite{mitchell2023detectgpt}. They need no cooperation from the provider, which is their appeal, but they have no key and no embedded signal to look for, so their accuracy is limited by how distinguishable machine text is from human text in general. OpenAI withdrew its own classifier of this kind in 2023, citing low accuracy~\cite{openai2023classifier}, and such detectors can be evaded by paraphrasing or by prompting a model to change its style~\cite{sadasivan2023can, liang2023biased}. Commercial detectors of this kind remain in use and appeared in coverage of the Claude announcement~\cite{techcrunch2026}. A watermark detector answers a narrower question, whether the signal tied to one specific key is present, and its false-positive rate is set by a threshold on its score rather than by how separable the two populations are (Section~\ref{sec:literature:measured}). Part of the confusion in the August debate comes from importing the failure record of one kind of tool into the discussion of the other. Figure~\ref{fig:mechanism} summarizes the mechanism and the distinction.

\subsection{What Has Been Measured}\label{sec:literature:measured}

The quality rows of Table~\ref{tab:claims} (Claim 1, Assurance 2), together with Assurance 1 on imperceptibility, are the most thoroughly measured. For both families, the measured cost is small with a modest increase in perplexity for distortion-based schemes~\cite{kirchenbauer2023watermark}, and for distortion-free schemes, perplexity and human evaluation at parity, as the construction guarantees~\cite{dathathri2024scalable}. Google's live comparison in Gemini found no difference in user feedback~\cite{dathathri2024scalable}; the comparison was run by the vendor, reported only in aggregate, and confined to the non-distortionary configuration, for which parity holds by construction. Nearly all of this evidence concerns prose. A study that used code instead found that a distortion-based watermark applied at every token lowers functional correctness, the fraction of generated programs that pass their reference tests, and that restricting the watermark to high-entropy tokens largely recovers it~\cite{lee2024sweet}. Claim 1 is therefore a result from the first family applied to a deployment that reports using the second and whether it carries over depends on the configuration and entropy cutoff in use.

The same rows concerns detectability rather than quality as a second measured result. Since the detector's score accumulates over tokens, detection strengthens with the length of the passage and with the entropy of each step, and both families report weaker detection on short or highly predictable text~\cite{kirchenbauer2023watermark, kirchenbauer2024reliability, dathathri2024scalable, he2026optimal}. The code study measured this directly, where entropy is low and detection degrades~\cite{lee2024sweet}. The users' code objection, restated in these terms, is that watermarked code may go undetected, which is a limitation for the provider, whose obligation is detectability, rather than a cost to the user. Section~\ref{sec:eval} measures correctness and detectability side by side on the same code outputs.

On removability (Claims 3 and 4, Assurance 5), the measured results place the boundary where Assurance 5 places it, at a full rewrite. Token-level edits and the mixing of watermarked spans into human-written text leave the watermark detectable as long as enough watermarked tokens remain~\cite{kirchenbauer2024reliability, kuditipudi2023robust}. Paraphrasing with a second model substantially reduces the per-token signal~\cite{krishna2023paraphrasing, sadasivan2023can}, and the watermark does not survive translation~\cite{he2024can}. Because detection strengthens with length, what counts as removal depends on the length of the passage and on the detector's threshold, and neither Claim 3 nor Claim 4 specifies either. A theoretical result goes further where an attacker able to judge the quality of outputs and to generate rewrites can remove any watermark that preserves quality~\cite{zhang2023watermarks}. The result depends on those assumptions and it bounds Claim 4 without establishing Claim 3.

The misattribution rows (Limitation 1, Concern 1) are where the literature is thinnest. The detector's threshold fixes a false-positive rate that is computed under the assumption that the text carries no watermark at all~\cite{kirchenbauer2023watermark, kuditipudi2023robust}. Text that a person wrote and a model then edited, the case Limitation 1 describes, is neither unwatermarked nor fully watermarked, and published error rates on such mixed text are sparse. Google's released detector returns one of three states, watermarked, unwatermarked, or uncertain, set by two thresholds~\cite{dathathri2024scalable, synthid2024docs}, an acknowledgment in the vendor's own tooling that the decision is not binary. No deployment has disclosed how its detector is calibrated or reported error rates at population scale, the quantities on which Concern 1 turns.

The disparate-impact row (Concern 2) rests on evidence about the other kind of tool. The finding that detectors flag non-native English writers at elevated rates concerns post-hoc classifiers, which flag text that a model finds predictable, and a constrained vocabulary reads as predictable~\cite{liang2023biased}. A watermark detector scores the presence of a key rather than a style, so the finding does not transfer directly; what could transfer is uneven detectability across languages, and the evidence there is limited to a few studies. The signal is not consistent under translation~\cite{he2024can}, and evaluations across 11 languages find that neither quality nor detectability is uniform across them~\cite{nemecek2026auditing, nemecek2026gets}.

\subsection{What Remains Open}\label{sec:literature:open}

The measured results stop at three points. The first is multilingual behavior at deployment scale: the family-level evidence is confined to open-weight models and no deployment has reported quality or detectability disaggregated by language or user group, which Table~\ref{tab:claims} lists as the settling evidence for Concern 2. The second is the absence of an agreed evaluation protocol. Benchmarks and toolkits exist~\cite{piet2025markmywords, tu2024waterbench, pan2024markllm}, but they differ in their metrics, text lengths, attack suites, and the false-positive rates at which detection is reported, so results for different schemes are not comparable with one another, let alone with a vendor's assurance; arguments have been made that watermarking without such standards does not constitute governance~\cite{nemecek2025watermarking}.

The third is coordination across providers. The Code of Practice has roughly 190 signatories, and each provider that watermarks does so under its own scheme and keys, with no shared detection interface. How a text that has passed through more than one provider's model is to be checked is an open question, and a text watermark and the signed provenance metadata that can accompany a generated file need not stay in step once the file is edited~\cite{nemecek2026provenance}. None of the three is a question about the mechanism; all three concern deployments and institutions (Sections~\ref{sec:problem} and~\ref{sec:moving_forward}).

\section{Is the Skepticism Proportionate?}\label{sec:eval}
Table~\ref{tab:claims} names the settling evidence for its quality rows: matched watermark-on and watermark-off outputs under a fixed quality protocol. No deployment has released such outputs nor can the watermark be disabled, so the comparison cannot be run on the system the August debate concerns. It can be run on the family. We therefore evaluate the open-source implementation of SynthID-Text~\cite{synthidtext2024code} through the MarkLLM toolkit~\cite{pan2024markllm} on two open-weight models in their instruction-tuned releases, Gemma-2-9B~\cite{gemmateam2024gemma} and Llama-3.1-8B~\cite{grattafiori2024llama}, in the non-distortionary configuration that vendors report deploying~\cite{dathathri2024scalable}. We furthermore take the same deployed interfaces, returning only a binary verdict~\cite{nemecek2025watermarking}.

For each model, we generate one watermarked and one unwatermarked response from 500 open-ended question prompts~\cite{DatabricksBlog2023DollyV2} from the same sampling seed, and a second unwatermarked response from a different seed. The first pair isolates the watermark; the second measures how far two ordinary runs drift apart, and the watermark's effect is read against that control. The same three arms were generated for 364 programming problems drawn from two standard code benchmarks~\cite{chen2021evaluating, austin2021program}, ten samples per problem. Sampling settings were identical in every arm and prompts on which any arm returned an empty response or a refusal were dropped from all arms (four of 500 on one model). The bounds that would count as equivalence were fixed before the runs: the change in perplexity under a scoring model from a third model family~\cite{qwen2024qwen25}, no more than 5\% in either direction; the win rate in a blinded pairwise preference, judged by an instruction-tuned model of the same family, each pair judged twice, once in each order, between 0.45 and 0.55; and the change in functional correctness (pass@1), within five percentage points.

On prose, every bound held on both models illustrated in Table~\ref{tab:eval}. The mean change in perplexity was $-0.3$\% on each model, and the confidence intervals are well inside the 5\% band; on Llama the seed control drifted $+1.5$\%, farther than the watermark itself. The judge returned a tie on most pairs and an overall win rate of 0.501 and 0.515 for the watermarked arm, and the semantic similarity between watermarked and unwatermarked responses, measured with a sentence-embedding model~\cite{xiao2023cpack}, matched the similarity between the two unwatermarked runs to within one half of one percent. On these models, the measured effect of the watermark on prose does not exceed the effect of changing the sampling seed.

\begin{table*}[!t]
\centering
\small
\begin{tabular}{@{}llcc@{}}
\toprule
 & & \textbf{Gemma-2-9B} & \textbf{Llama-3.1-8B} \\
\midrule
\multirow{5}{*}{\shortstack[l]{Prose quality\\500 prompts}}
 & Perplexity change (\%) & $-0.3$ $[-2.2, +1.6]$ & $-0.3$ $[-2.7, +2.3]$ \\
 & \quad Seed control & $+0.0$ $[-1.9, +2.0]$ & $+1.5$ $[-1.3, +4.3]$ \\
 & Judge win rate & $0.501$ $[0.484, 0.518]$ & $0.515$ $[0.490, 0.540]$ \\
 & Similarity, watermark pair & $0.946$ $[0.942, 0.950]$ & $0.927$ $[0.921, 0.933]$ \\
 & \quad Seed control pair & $0.945$ $[0.941, 0.949]$ & $0.928$ $[0.922, 0.934]$ \\
\midrule
\multirow{4}{*}{\shortstack[l]{Code correctness\\364 problems}}
 & pass@1, watermarked (\%) & $61.8$ & $60.7$ \\
 & pass@1, unwatermarked (\%) & $61.4$ & $63.8$ \\
 & Difference (points) & $+0.4$ $[-1.4, +2.1]$ & $-3.1$ $[-6.0, -0.3]$ \\
 & \quad Seed control & $-0.6$ $[-1.3, +0.1]$ & $+0.6$ $[-0.7, +1.8]$ \\
\midrule
\multirow{4}{*}{\shortstack[l]{Detection\\1\% false-positive rate}}
 & Prose TPR, 200 tokens (\%) & $39$ & $39$ \\
 & Prose TPR, 400 tokens (\%) & $59$ & $56$ \\
 & Prose AUROC, full length & $0.78$ & $0.82$ \\
 & Code AUROC, full length & $0.55$ & $0.57$ \\
\bottomrule
\end{tabular}
\vspace{1em}
\caption{Family-level evaluation of SynthID-Text in its non-distortionary configuration. Brackets give 95\% confidence intervals. Each seed-control row compares two unwatermarked generations that differ only in sampling seed. Detection thresholds are set at each length to a 1\% false-positive rate on the unwatermarked outputs.}
\label{tab:eval}
\end{table*}

For the coding paradigm, on Gemma, correctness was equivalent: 61.8\% of watermarked samples passed their reference tests against 61.4\% unwatermarked, a difference of $+0.4$ points with the control at $-0.6$. On Llama, the watermarked rate was 3.1 points lower, 60.7\% against 63.8\%; the confidence interval excludes zero and extends just past the five-point bound, and the control shows no comparable movement, so the cost is present but small for this model configuration. Both results show a distortion-based watermark applied at every token lowered correctness outright~\cite{lee2024sweet}, and the non-distortionary configuration measured here reduces that cost to a few points on one model and below measurement on the other.

\begin{figure}[!t]
\centering
\includegraphics[width=\columnwidth]{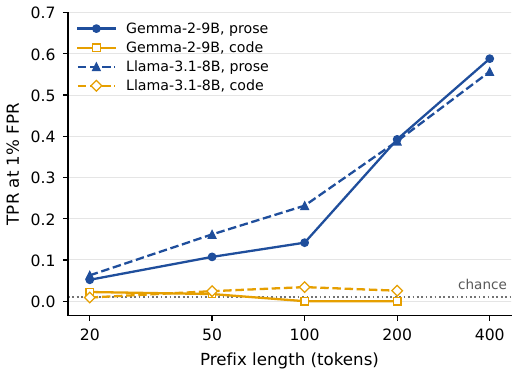}
\caption{Detection as a function of passage length for the outputs of Table~\ref{tab:eval}. Each curve reports the true-positive rate of the mean g-value detector at a false-positive rate of 1\%, with the threshold re-estimated at each length from the pooled unwatermarked outputs. Prose detection strengthens with length on both models; code detection stays near chance at every length, and few code outputs are long enough to test further.}
\label{fig:detection}
\end{figure}

The same outputs were then scored for detection with the mean g-value detector of Section~\ref{sec:literature:families}. The threshold was set on the pooled unwatermarked outputs to a false-positive rate of 1\% and re-estimated at each passage length, so that short and long passages are compared at the same false-positive rate. As the literature predicts, detection strengthens with length, reaching a true-positive rate of 39\% at 200 tokens and 56\% to 59\% at 400 tokens on the two models, and the area under the receiver operating characteristic curve, a threshold-free summary of separation for which 0.5 is chance, reaches 0.78 and 0.82. Code does not. The same detector, with the same keys, stays at 0.55 and 0.57 at every length the code outputs attain. Since the paired arms share a sampling seed, the watermark leaves a generation untouched and between 25\% and 31\% of code samples were identical, character for character, with and without the watermark. The rest of the samples were short and fewer than one code sample in a hundred reached 200 tokens on Gemma. Code is disadvantaged twice, by the entropy of each step and by the length of the passage, and Figure~\ref{fig:detection} shows the two domains separating as length grows.

Comparing Section~\ref{sec:conflict}, the loudest claim of the August debate, that watermarking unavoidably degrades generated code, finds at most a three-point effect on one model and none on the other, under the configuration the vendors report using; at the family level, the quality skepticism is largely disproportionate. Detection on code barely rises above chance. That is the restatement Section~\ref{sec:literature:measured} gave the code objection, now measured directly: on low-entropy content, the watermark's cost falls not on the quality the user receives but on the detectability the provider owes~\cite{he2026optimal}.

These results concern the family: two open-weight models, the public implementation, the default non-distortionary configuration, English prompts, a model rather than human raters in the preference test, and a detector that scores by the raw mean where the reference implementation also distributes a trained detector~\cite{synthidtext2024code}. None of this measures the deployment. Which configuration Anthropic runs, at what strength, with what entropy cutoff, and under what calibration cannot be tested from outside, so the numbers above bound the users' claims about the method while leaving the vendor's assurances about the product where Section~\ref{sec:conflict} left them, resting on access that no one outside the company holds. The design is not the obstacle; the same three arms run on the deployed system would settle the quality rows of Table~\ref{tab:claims} directly, and Section~\ref{sec:problem} takes up why that evidence does not exist.

\section{What Cannot Be Verified}\label{sec:problem}

Section~\ref{sec:eval} ran the three arms that would settle the quality rows of Table~\ref{tab:claims}, but only for the family; no public tool can run them on a deployment. The gaps are of two kinds: (1) withheld access and (2) institutional. Withheld access involves withholding matched outputs with a detector that reports a score and the deployed configuration, both of which the vendor could supply. The institutional gaps are key management, accountability for detection, and evaluation itself are beyond any one vendor's ability to prove, because the bodies that verification requires, an agreed protocol, an independent auditor, and a means of coordination across providers, do not exist. This would mean an agreed protocol, an independent auditor, and a means of coordination across providers which is beyond scope.

For what is withheld, the category (b) rows wait on access, as Section~\ref{sec:conflict} established. However, what was not stated is the price of supplying said access. Matched watermark-on and watermark-off outputs are produced by a configuration flag and a detector that reports a score is a product decision. The detectors that currently exist compute a score internally and the public interfaces reduce it to a verdict (e.g., watermarked or not watermarked)~\cite{dathathri2024scalable, synthid2024docs}. Google's live comparison shows that a vendor can produce such evidence at production scale and also what the vendor-run version can't show, since no one outside the company could validate it (Section~\ref{sec:literature:measured}). The flagging rate on human text edited by the model and the measurement Limitation 1 requires are comparable. None of this evidence is expensive to produce, so its absence reflects a decision rather than a limitation.

The second withheld item is the configuration. Section~\ref{sec:literature:families} deferred the question of an entropy cutoff, a threshold below which tokens are left unwatermarked. Watermarking every token of code costs correctness in the distortion-based family~\cite{lee2024sweet}, the non-distortionary configuration leaves code detection near chance (Section~\ref{sec:eval}), and the vendor itself states that code carries less watermark~\cite{anthropic2026watermark}. A deployment therefore has every reason to switch the watermark off where entropy is low, and no deployment discloses whether it does, at what threshold, or for what share of its output. The result is a transparency mechanism that may be silently inactive where the outputs are least likely to carry the watermark. Code and other constrained formats, fall under the same detectability obligation as everything else~\cite{euaiact2024}, and whether that obligation is met in substance cannot be determined from outside.

The traceability rows cannot be resolved by what a vendor might release about outputs. Section~\ref{sec:conflict} left Claim 2 and Assurance 4 as mirror images that no external test can distinguish because both are claims about infrastructure (i.e., how many keys exist, whether they are assigned per customer or shared, how they are rotated, who holds them). Key management is a subject of audit in other security-critical settings~\cite{barker2020keymanagement}; here no auditor has access, so the strongest available answer to Claim 2 remains a denial. The planned detection API extends the problem from the keys to the verdicts~\cite{techcrunch2026}. The vendor will calibrate the detector, decide who may query it, and adjudicate disputes over its output, so a person whose text is wrongly attributed to the model, the case of Limitation 1 and Concern 1, can challenge the finding only through the tooling of the party that produced it.

The remaining absences would survive full cooperation. A vendor that releases matched outputs, a scored detector, and its configuration would still have no agreed protocol under which the release resolves anything, since the available benchmarks differ in metrics, passage lengths, attack suites, and operating points, and the same release could pass one and fail another (Section~\ref{sec:literature:open}). There is currently no body to conduct or accredit an audit, and the disaggregated reporting that Concern 2 waits on exists for no deployment. Coordination is in a similar state, as each signatory that watermarks, does so under its own scheme and keys, no shared detection interface exists, and the signed provenance metadata that can accompany a generated file can come to contradict the text watermark once the file is edited~\cite{nemecek2026provenance}. The Code of Practice commits its signatories to detection systems and specifies none of the above~\cite{codeofpractice2026}.

What follows from these absences together is the August reception. Section~\ref{sec:eval} located the error in the users' claims; the cause lies here, in a deployment about which nothing could be checked. A public offered assurances with no means of testing them will reason from the worst case available, and the contradictory objections, the unvalidated removal tools, and the cancellations are what such a system produces, not what public misunderstanding produces. Each absence named is a resource some identifiable party could supply.

\section{Moving Forward}\label{sec:moving_forward}

Table~\ref{tab:claims} names the evidence that would settle each contested assertion and Section~\ref{sec:problem} traced why that evidence is missing. Each missing item needs a requirement assigned in a party in a position to supply it. We formalize five in total.

First is the release via the vendor. For example, matched watermark-on and watermark-off outputs, together with a detector that reports its score and the operating points at which a score becomes a verdict. Together these requirements move the category (b) rows of Table~\ref{tab:claims} into category (a): the quality and perceptibility assurances become testable on the deployment rather than the family, the removability assertions can be run against the deployed detector, and the rate at which human text edited by the model is flagged can be measured outside the vendor.

The second requirement is disclosure of the deployed configuration such as which of the configurations of Section~\ref{sec:literature:families} is running and at what strength, whether an entropy cutoff is in use, and, if one is, its threshold and the share of output, by content type, generated below it. These parameters reveal nothing about the key and their publication does not ease removal, because the attacks that succeed make no use of them (Section~\ref{sec:literature:measured}). Publication turns an assurance about coverage into a statement that outside parties can check.

Thirdly is the audit requirement. Audits are required because the traceability pair cannot be settled by anything a vendor releases about outputs. With key management acting as the subject and access as the vendor's to grant, the missing element is a body accredited to receive it. Detection needs its own governance via a documented access model, disclosed calibration with error-rate guarantees at stated operating points, error rates reported at population scale, and a means of disputing an attribution that does not run through the vendor alone. Publicly detectable schemes allow anyone holding a public key to verify the watermark without being able to forge it~\cite{fairoze2023publicly}, and intermediate designs can grant scored access to accredited parties. The audit requirement is one that is disclosed with stated guarantees.

The fourth requirement is the protocol under which any of these releases is read. Results resolve little, if any, while benchmarks disagree on the quantities listed in Section~\ref{sec:literature:open}. There is a standard mix of benchmarks~\cite{piet2025markmywords, tu2024waterbench, pan2024markllm}, but they must require reporting disaggregated by language and content type~\cite{nemecek2026gets}, the settling evidence Table~\ref{tab:claims} names for Concern 2. NIST has published guidance on synthetic-content transparency that covers watermarking and its evaluation~\cite{nist2024synthetic}, and Article 50(7) empowers the Commission to make common rules on marking and detection binding where the code of practice proves inadequate~\cite{euaiact2024}. Neither route has yet produced a protocol specific enough to run.

The fifth and final requirement is collective as each provider watermarks under its own scheme and keys, and neither needs to be shared, however the interface does. A common way to submit a text and receive a scored answer at stated error rates from each provider's detector is the precondition for checking text that has passed through more than one model (Section~\ref{sec:literature:open} left open). The Code of Practice commits signatories to detection systems and is the venue in a position to specify the interface~\cite{codeofpractice2026}. The same specification should state how the text watermark and the signed provenance metadata accompanying a generated file are to remain consistent under editing, or how their divergence is to be detected~\cite{nemecek2026provenance}. 

Of the five requirements, the first two are within a single vendor's power; the last three require an auditor, a protocol, and a venue that no vendor can supply for itself.

\section{Outlook}\label{sec:outlook}

Every generative AI provider (serving the European Market) is now bound by the Article 50 obligations which took effect August 2026~\cite{euaiact2024}. Roughly 190 signatories of the Code of Practice have committed to providing detection systems~\cite{codeofpractice2026}. This work is not specific to one company as Google deployed the same watermarking two years earlier (2024)~\cite{dathathri2024scalable} to far less public attention compared to Anthropic's deployment~\cite{anthropic2026watermark, anthropic2026marking} which, at the time of writing, is the most recent and most public. Additional companies, such as OpenAI, have already built a text watermark and withheld it~\cite{openai2024provenance, wsj2024openaiwatermark}. The providers that have yet to deploy face the same obligations with the same missing verification aspects as well as the same reception documented regarding our claims, assurances, and no independent way of weighing one against the other.

There are two broad avenues to explore: (1) watermarking matures into trusted provenance infrastructure, allowing evaluation under common protocols and auditing by parties other than their operators and (2) watermarking moves to being a compliance ritual~\cite{nemecek2025watermarking} because the law requires said marking and each new deployment replays the August reception at larger scale. The avenue that arrives won't be decided by the watermarking schemes themselves since they are not the limiting factor. The limitations instead involve a quality cost that is negligible to small, and the limits that do exist, for instance on low-entropy content~\cite{dathathri2024scalable, lee2024sweet} and under rewriting~\cite{krishna2023paraphrasing, sadasivan2023can}, are not hidden. What will decide the avenue is aspects surrounding the watermarking scheme: the access, the disclosure, and the institutions described in Sections~\ref{sec:problem} and~\ref{sec:moving_forward}.

The research literature already supplies evaluation protocols, benchmarks, and toolkits for quality, detectability, and robustness~\cite{piet2025markmywords, tu2024waterbench, pan2024markllm}. Section~\ref{sec:eval} demonstrated such on a small scale that a family-level comparison needs little more than an open-weight model, public tooling~\cite{synthidtext2024code}, and a configuration flag. Article 50, while it requires generated content to be marked and detectable, does not require the mark be evaluated under a shared protocol, audited by anyone independent of the vendor, or detectable through interfaces that the vendors do not themselves govern~\cite{euaiact2024}. These verification aspects must arrive while there is still public trust left for it to protect.

\bibliographystyle{ieeetr}
\bibliography{references}

% \EOD

\end{document}